\documentclass[prb,twocolumn]{revtex4}

\usepackage{graphicx,amsmath}

\begin{document}

\title{Multi-scale Extensions to Quantum Cluster Methods for Strongly Correlated Electron Systems}
\author{
C.\ Slezak$^{1}$,
M.\ Jarrell$^{1}$, 
Th.\ Maier$^{2}$ and
J.\ Deisz$^{3}$}

\address{
$^{1}$University of Cincinnati, Cincinnati, Ohio, 45221, USA \\
$^{2}$Oak Ridge National Laboratory, Oak Ridge, Tennessee, 37831, USA \\
$^{3}$University of Northern Iowa, Cedar Falls, Iowa, 50614, USA}

\date{\today}

\begin{abstract}

  A numerically implementable Multi-scale Many-Body approach to
  strongly correlated electron systems is introduced. An extension to
  quantum cluster methods, it approximates correlations on any given
  length-scale commensurate with the strength of the correlations on
  the respective scale. Short length-scales are treated
  explicitly, long ones are addressed at a dynamical mean-field level and 
  intermediate length-regime correlations are assumed to be
  weak and are approximated diagrammatically. To illustrate and 
  test this method, we apply it to the one dimensional Hubbard model.
  The resulting multi-scale self-energy provides a very good 
  quantitative agreement with substantially more numerically expensive, 
  explicit Quantum Monte-Carlo  calculations. 

\end{abstract}
\maketitle

\section{Introduction}

Strongly correlated electron systems are characterized by a manifold
of complex, competing phenomena, which emerge in the thermodynamic
limit. The underlying mechanisms involve correlations on all length
scales. Currently no single feasible numerical method exists to
accurately address these correlations at all lengths. Both finite size
and mean-field calculations alike are faced with these limitations.
However, interest in areas such as quantum phase transitions and 
magnetically-driven superconductivity point to the need for
numerical schemes that accurately bridge the short and long
length scales. 

Quantum cluster techniques\cite{maier2}, constitute a good
staring point in addressing the entire range of correlations by
dividing the problem into two length regimes; explicitly solving for
short ranged correlations and approximately for the remaining longer
length-scales. One such method, the dynamical cluster approximation
(DCA)\cite{dca}, maps the lattice problem onto an embedded cluster
problem.  In doing so, short-ranged correlations within a given cluster
are treated accurately, while the remaining longer-ranged correlations
are approximated on a dynamical mean-field level.  However, the extent of
length-scales which can thus be accurately addressed beyond the mean-field level 
is severely limited by the numerical expense involved.

This limitation results in an inadequate treatment of medium ranged
correlations which are outside the scope of explicit calculations. We
introduce a Multi-scale Many-Body (MSMB) approach which addresses each
length-scale using approximations adequate for the strength of the
correlations on the respective scale.  The strongest, local and short
ranged correlations are well accounted for in traditional, numerically
exact implementations of the DCA. Correlations, except for in the vicinity 
of phase transitions, fall off rapidly with distance and 
are hence considered weaker in the intermediate length-regime. 
However, these correlations remain significant and
will hence be approximated diagrammatically. Only the remaining third
regime of the longest length scale will be treated at the dynamical mean-field
level.

The perturbative inclusion of correlations on an intermediate 
length-scale within
a multi-scale approach has previously been explored by Hague 
{\it et.~al.}~\cite{hague}. Contributions to the single-particle 
self-energy on various scales were linked in a hybrid approach. 
However, the inherent perturbative nature of the approach limited it
to high temperatures and/or weak coupling strengths.
In this work, we present a non-perturbative two-particle diagrammatic
approach to the intermediate length-regime.

To illustrate and test this scheme, this
MSMB is applied to the one dimensional Hubbard
model\cite{hubbard}.  While a formally simple model, the Hubbard model
contains a multitude of the underlying physics of correlated electron
systems. It therefore lends itself ideally as a benchmark for the
method.  We show that the MSMB approach yields results in very good
quantitative agreement with explicit large cluster calculations.

Before we proceed to provide detailed results in section~\ref{results} 
we first establish the theoretical basis for the method.
A subsequent outlook on further developments in MSMB techniques is
provided in section~\ref{outlook}, concluding with a brief summary.

\section{Formalism}
\label{formalism}

For simplicity, we will use the one dimensional Hubbard model to illustrate the
MSMB formalism. This low dimension is also the most difficult regime for quantum cluster
approaches like the DCA.
The Hubbard Hamiltonian is given by

\begin{equation}
\begin{split}
  H=-\sum_{<ij>} t(c^\dagger_{i\sigma}c_{j\sigma}+H.c.)+\epsilon \sum_{i\sigma}n_{i\sigma} \\
  + U\sum_i (n_{i\uparrow}-1/2)(n_{i\downarrow}-1/2)
\end{split}
\end{equation}

\noindent with $c^{\dagger}_{i\sigma}$ creating an electron of spin
$\sigma$ at site $i$ and local density
$n_{i\sigma}=c^{\dagger}_{i\sigma}c_{i\sigma}$. The first part, the
kinetic term, allows hopping between adjacent lattice sites with
transfer integral $t$.  The second term is the on-site Coulomb
repulsion making a doubly occupied lattice site unfavorable. 
Throughout the remainder of this paper we
choose the bare bandwidth $W=4t$ as the unit of energy by setting
$t=0.25$ and work at fixed filling $n=0.75$.

\subsection{DCA} 
\label{secdca}

The DCA is a systematic quantum cluster theory that maps the lattice
problem onto a self-consistently embedded cluster problem. It is an
extension of the dynamical mean-field theory
(DMFT)\cite{muller,metzner} which systematically incorporates
non-local correlations.  In the limit when the cluster size is one
(i.e.\ single site), it recovers the purely local DMFT solution,
systematically incorporates non-local corrections
as the cluster size increases, and
finally becomes exact when the cluster size equals the size of
the lattice.

The respective approximations for the DMFT and DCA may be derived by
approximating the Laue function which describes momentum conservation
at the vertices of the irreducible diagrams:

\begin{equation}
\label{eq:laue}
\Delta_{(k_1,k_2,k_3,k_4)}\equiv\sum_{r} e^{\imath r(k_1+k_2-k_3-k_4)}
\end{equation}

\begin{equation}
\label{eq:dmft}
\Delta_{exact}= N\delta_{k_1+k_2,k_3+k_4}~.
\end{equation}

\begin{figure}
\centerline{\includegraphics*[width=2.3in]{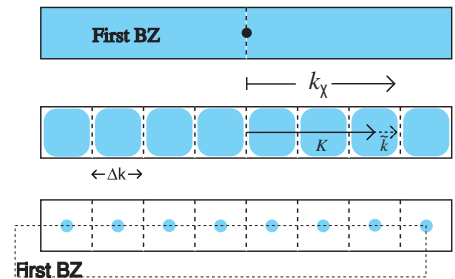}}
\caption{In the DMFA all electron propagators are averaged over the
  entire Brillouin Zone (BZ); effectively mapping the lattice onto a
  single point ({\it top}). In the DCA, we break the BZ into several
  sub-cells which are now in turn averaged over mapping the lattice
  onto a finite sized cluster ({\it bottom}).}
\label{fig:dca}
\end{figure}

\noindent In the DMFT, the Laue function is
approximated with $\Delta_{DMFT}=1$ for all combinations
of $k_1$, $k_2$, $k_3$ and $k_4$. In doing so, all electron
propagators in the self-energy diagrams may be averaged over the first
Brillouin Zone (BZ); thus relinquishing any momentum dependence of the
self-energy. Hence, the DMFT lattice Green's function contains local
correlations of the system but is unable to capture non-local
correlations.  The DCA sets out to systematically include these
non-local contributions. This is accomplished by partially restoring
momentum conservation of the irreducible vertices.  We divide the BZ
into $N_c$ identical discrete sub-cells as illustrated in
Fig.~\ref{fig:dca}. The center of each cell is labeled by $K$, and the
surrounding points by $\tilde{k}$, so that any arbitrary
$k=K+\tilde{k}$.  In the DCA, this partial momentum conservation is
expressed by the Laue function:

\begin{equation}
\Delta_{DCA}=N_c\delta_{K_1+K_2,K_3+K_4}~.
\end{equation}

\noindent Therefore, all propagators may freely be summed over
intra-cell momenta $\tilde{k}$, yielding the coarse grained Green's
function

\begin{equation}
\label{coarg}
\overline {G}(K,i\omega_n)=\frac{N_c}{N}\sum_{-\frac{\Delta k}{2}<\tilde{k}\leq
\frac{\Delta k}{2}}G(K+\tilde{k},i\omega_n)~~.
\end{equation}
 
\noindent In so doing, only momentum conservation of magnitude $\Delta
k<(2\pi/N_c)$ is neglected, while larger inter-cell transfers are
preserved.  The resulting self-energy diagrams are now those of a
finite cluster of size $N_c$ where each lattice propagator has been
replaced by its coarse-grained analog, and the remaining cluster
problem is defined by $\overline{G}(K,i\omega_n)$. We can write for
the DCA lattice Green's function

\begin{equation}
G(K+\tilde{k},i\omega_n)=\frac{1}{i\omega_n+\mu-\epsilon(K+\tilde{k})-
\Sigma(M(K+\tilde{k}),i\omega_n)}
\end{equation}

\noindent where $M(k)$ is a function which maps momentum $k$ residing
in a certain sub-cell of the BZ to its cluster momentum $K$ and the
lattice self-energy is approximated by that of the cluster problem.

\begin{figure}
\centerline{\includegraphics*[width=2.3in]{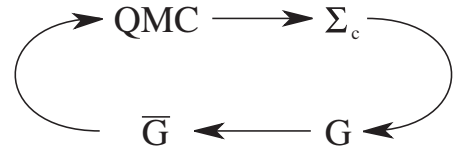}}
\caption{Self-consistency loop for the DCA.} 
\label{fig:dcasc}
\end{figure}

The remaining embedded cluster problem must be solved
with a self-consistency requirement that the
Green's function calculated on the cluster $G_c(K,i\omega_n)=\overline
{G}(K,i\omega_n)$. Fig.~\ref{fig:dcasc} depicts the corresponding DCA
algorithm: Starting with an initial guess for the self-energy, we
construct the coarse-grained Green's function $\overline {G}$ from the
corresponding lattice $G$ (see Eq.~\ref{coarg}). In the next step, we
utilize one of the many available cluster solvers to determine the
cluster self-energy. This is the numerically most involved step and a
variety of numerical techniques may be applied.  At this point, we use
the new estimate for the self-energy to re-initialize the
self-consistency loop. It is important to notice that in this
procedure only the irreducible lattice quantities are approximated by
their cluster equivalent - i.e.\ the self-energy.

The DCA has been successfully implemented with a variety of cluster
solvers of which some are exact but limited in cluster size, while
others are applicable up to larger length-scales but involve varying
degrees of approximation. Some of the cluster solvers which have been
used in conjunction with the DCA include the non-crossing
approximation (NCA)\cite{maier}, the fluctuation exchange
approximation (FLEX)\cite{aryanpour,hague} and the Quantum Monte Carlo
(QMC)\cite{dca} method. While NCA and FLEX involve various levels of approximations,
QMC is of special interest since it provides an
essentially numerically exact solution to the problem.
Although the QMC constitutes a precise cluster solver, it becomes
prohibitively expensive for large clusters. The range of applicability
of exact calculations is thus restricted to relatively short
length-scales. However, various properties of strongly-correlated
systems are not accounted for (e.g.~Mermin-Wagner
theorem\cite{mermin}) due to the absence of long-ranged fluctuations
in these solutions.

\subsection{Multi Scale Method}
\label{multiscale}

The inability of a single solver within the DCA to numerically address
long-ranged correlations explicitly, motivates a MSMB approach
where the problem is divided further, incorporating a third,
intermediate length-regime.  In this approach, the lattice problem is
mapped onto two clusters of different size each of which contributes
correlations of length-scales up to the linear extent of their
respective cluster size. The respective cluster problems are addressed
using approximations adequate with the strength of the correlations on
the respective scale.  We choose a small DCA cluster of size $N_c^{(1)}$
to be solved using the QMC, thus explicitly accounting for the
shortest ranged correlations in the system. Next, we invoke a second,
larger cluster of size $N_c^{(2)}$ to address the intermediate length
regime. Except in the vicinity of phase transitions, correlations on 
these longer length-scales are weaker and the corresponding self-energy 
is approximated diagrammatically.

\begin{figure}
\centerline{\includegraphics*[width=2.3in]{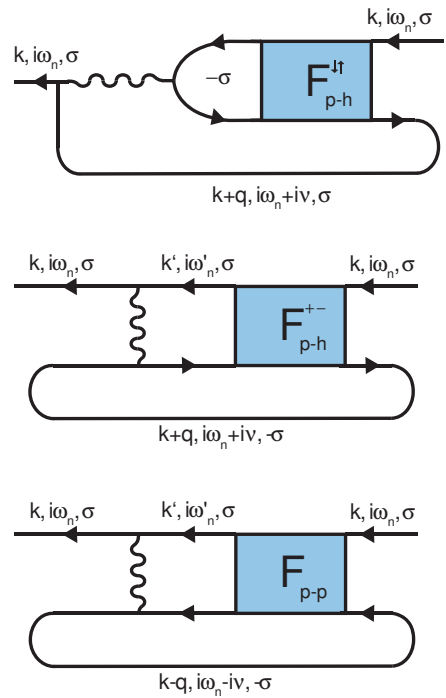}}
\caption{Diagrams relating the self-energy to the reducible
  two-particle longitudinal spin and charge ({\it top}) and transverse
  spin ({\it middle}) vertices. A similar relation is obtained for the
  particle-particle channel ({\it bottom}).}
\label{fig:Fig2}
\end{figure}

\begin{figure}
\centerline{\includegraphics*[width=3.3in]{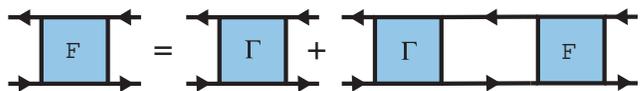}}
\caption{ Bethe-Salpeter equation relating the reducible two-particle vertex $F$ to the 
irreducible vertex function $\Gamma$.}
\label{fig:bse}
\end{figure}

We build a suitable approximation by considering the single-particle
self-energy (the Hartree term is not explicitly shown) written in
terms of the reducible vertex $F$ in both the particle-hole and
particle-particle channel as depicted in Fig.~\ref{fig:Fig2}.  
$F$ in turn is related to the irreducible vertex $\Gamma$ via the Bethe-Salpeter equation (see
Fig.~\ref{fig:bse})

\begin{equation}
\label{eq:feqg}
\begin{split}
F(k,k',q&;\imath \omega_n,\imath \omega_n', \imath \nu) = 
\Gamma(k,k',q; \imath \omega_n, \imath \omega_n',\imath \nu)\\
&+ \Gamma(k,k'',q; \imath \omega_n, \imath \omega_n'',\imath \nu) 
\times \chi^0(k'',q; \imath \omega _n, \imath \nu)\\
&\times F(k'',k',q;\imath \omega_n'',\imath \omega_n', \imath \nu)
\end{split}
\end{equation}

In perturbation theory, the approximations to the self-energy are
often made at the level of the irreducible vertex function. In the
simple approximation $\Gamma=U$, the resulting self-energy diagrams in
Fig.~\ref{fig:Fig2} are those of the FLEX. The resulting method
provides qualitatively correct long-range properties, but short ranged
correlations are addressed inadequately - e.g.\ local moment
formation.

The failure of the FLEX, and similar perturbative approaches based on
the irreducible vertex, occurs in the stronger coupling regime. This
is a result of neglecting higher order correction terms which
re-normalize the vertex. In a non-perturbative approach, Y.M. Vilk
{\it et.al.}~in Ref.~\onlinecite{vilk} considered a renormalized
static vertex to address this problem.

In the MSMB method however, we introduce the following
non-perturbative approximation for the intermediate length-regime: The
large cluster irreducible two-particle vertex is approximated by the
small cluster irreducible vertex thus preserving the exact short
ranged correlation (on the small cluster) and approximating the
intermediate ranged ones (on the large cluster):

\begin{equation}
\label{eq:gclust}
\Gamma(K_2,K_2',Q_2;i\omega_n,i\omega_n',i\nu)\rightarrow \Gamma 
(K_1,K_1',Q_1;i\omega_n,i\omega_n',i\nu)
\end{equation}

\noindent where momenta labeled with subscript $1$ are cluster momenta
on the small cluster, while subscript $2$ denotes large cluster
momenta.  The approximated self-energy on the large cluster is
evaluated from the Dyson equation depicted in Fig.~\ref{fig:Fig2}. It
preserves all short ranged correlations and includes long-wave length
contributions which emerge from the inclusion of large cluster
corrections through the bare bubble $\chi^0$, i.e.\ the reducible part
of the vertex.  However, by implementing the DCA on a two-particle
irreducible level (i.e.\ replacing the large cluster irreducible
vertex with the small cluster equivalent), the resulting self-energy
is inherently different from the single-particle DCA self-energy that
one would obtain from the usual, direct DCA/QMC simulation of the
large cluster (see Sec.~\ref{secdca}).  Only in the case of infinite
dimensions, where the problem becomes purely local, are these two
approaches equivalent.

A few remarks about the above approximation are in order. 
The approximation in
Eq.~\ref{eq:gclust} breaks the crossing symmetry of $F$ (as it is related to
$\Gamma$ by Eq.~\ref{eq:feqg}), since $F$ now contains long-ranged
corrections beyond the linear extent of the small cluster 
only in the channel it is calculated in. Hence, the
self-energy contribution have to be evaluated in all three depicted
channels (see Fig.~\ref{fig:Fig2}). Furthermore, this %We want to emphasize that this
MSMB approach requires a full knowledge of the momentum and frequency
exchange on the cluster and hence the evaluation of $\Gamma$ on the
small cluster still involves extensive numerical calculations and is
limited by storage/memory requirements (see Sec.~\ref{outlook} for
details).

This difficulty of obtaining the small cluster irreducible vertex
$\Gamma$, necessitates a further long-ranged approximation for
$\Gamma$. In ref.~\onlinecite{agd} Abrikosov {\it et.~al.}~have, in
their study of Fermi liquid theory, identified a sub-class of
self-energy diagrams which convey the long length-scale properties of
a system. These are constructed from irreducible vertex functions
with zero external momentum and frequency transfer. In applying this
restriction we expect to capture long ranged correlations which are
characterized by small momentum transfers $q$.  The resulting
long wave-length approximated irreducible vertex, denoted by
the index $\lambda$, is given by

\begin{equation}
\label{eq:gamma}
\begin{split}
\Gamma(K_1,K'_1,Q_1&;i\omega_n,i\omega_n',i\nu_n) \rightarrow\\
&\Gamma^{\lambda}(K_1,K'_1,Q_1=0;i\omega_n,i\omega_n',i\nu_n=0)~~.
\end{split}
\end{equation}

The MSMB approximation at the level of the irreducible vertex is expected to provide the best
description for the most dominant two-particle processes. Furthermore, any 
phase transitions would manifest themselves in instabilities of the reducible vertex 
which contains the singular structure while the irreducible vertex would 
remain analytic throughout.
However, the MSMB result is expected to remain most valid away from phase transitions where intermediately ranged correlations are weak. When these correlations are not weak, a diagrammatic approximation of the self energy may contain significant errors, but such an approach can nonetheless represent non-trivial aspects of strong correlations such as non-Fermi-liquid behavior\cite{manske,jdeisz}.
Further details on the calculation of the $\lambda$ self-energy are provided in section~\ref{lcs}.

The self-energy obtained in this
$\lambda$-approximation only correctly accounts for long-ranged
fluctuations of the system. The remaining short length-scale
contributions have been neglected and have to be accounted for
separately as will be discussed in a subsequent section~\ref{ansatzsec}.

\subsection{A Conserving Approximation}

There are currently two widely used approaches which are sufficent to show an 
approximation is conserving. For one, G.~Baym (see Ref.~\onlinecite{baym1}) 
has shown that, with local conservation of spin, charge, momentum and energy
at each vertex, it is sufficient for the irreducible
vertex to be a functional derivative of the large cluster self-energy
$\Gamma(G(K,\omega),U)=\delta \Sigma(G(k,\omega),U)/\delta G(k,\omega)$.
However, within both the DMFT and DCA momentum conservation is partially violated
at each of the internal vertices as described by their respective Laue functions 
(Eq.~\ref{eq:laue}). This violates certain Ward identities and hence neither   
the DMFT nor the DCA constitutes a conserving approximation (see ref.~\onlinecite{hettler}
for details).

In an alternative approach, following the arguments of Baym and Kadanoff\cite{baym}
an approximation can also be shown to be conserving as long as it fulfills both the 
requirements of 1) the inversion symmetries of $F$ to be preserved and
2) that the two-particle correlation function tends to the single
particle Green's functions via the Dyson's equation. It is straightforward
to show that the approximation for the self-energy on the large
cluster satisfies these requirements. Thus, the MSMB 
{\it formalism} based on the explicit small cluster vertex 
(Eq.~\ref{eq:gclust} only) constitutes a conserving approximation for 
the large cluster as it restores momentum dependence of the large cluster 
self-energy. 

With the introduction of the $\lambda$-approximation (Eq.~\ref{eq:gamma}) short 
ranged correlations are neglected and have to be supplemented as 
discussed in the subsequent section. The resulting self-energy breaks the 
conservation laws with corrections of order $1/L_c^{(2)}$ (where $L_c^{(2)}$ is
the linear size of the larger cluster). It should be reiterated however that the necessity 
of the second (long wavelength) approximation is only temporary until the hurdle of large 
memory requirements can be met.

\subsection{Ansatz}
\label{ansatzsec}

To account for the omitted set of short length-scale self-energy
diagrams in the $\lambda$-approximation, we substitute appropriate
diagrams from the small cluster QMC result.  This diagrammatic
substitution between the different length-scales in the MSMB method is
done by means of an analytic {\it Ansatz}.

One possible implementation of the MSMB method which yields a
self-energy containing correlations on both long and short
length-scales is given by the real-space {\it Ansatz}:

\begin{equation}
\label{eq:rseq}
\begin{split}
  &\Sigma^{(N_c^{(2)})}(x_i,x_j)=\\
  &\left\{\begin{array}{cl}
      \Sigma^{(N_c^{(1)})}_{QMC}(x_i-x_j), & |x_i-x_j| \leq \frac{N_c^{(1)}}{2} \\
      \Sigma^{(N_c^{(2)})}_{\lambda}(x_i-x_j), & \frac{N_c^{(1)}}{2} <
      |x_i-x_j| \le
      \frac{N_c^{(2)}}{2} \\
      0, & otherwise
	   \end{array}\right.
\end{split}
\end{equation}

\noindent In this formalism, the MSMB self-energy is constructed by taking
all contributions of lengths up to the linear sizes of the small
cluster ($N_c^{(1)}/2$) from the exact QMC result and the remaining
longer ranged contributions are complemented by the $\lambda$-approximated
large cluster self-energy.  Note, the multi-scale self-energy obtained
in Eq.~\ref{eq:rseq} has lost its spatial continuity.
Fourier-transforming the individual length-scale contributions in
Eq.~\ref{eq:rseq} yields

\begin{equation}
\label{eq:realspace}
\begin{split}
  \Sigma^{(N_c^{(2)})}&(K_2,i\omega_n) = \Sigma^{(N_c^{(1)})}_{QMC}(x=0,i\omega_n)\\
  &+ \sum_{i=1,\frac{N_c^{(1)}}{2}}2~cos(i~K_2) \Sigma^{(N_c^{(1)})}_{QMC}(x=i,i\omega_n)\\
  &+ \sum_{j=\frac{N_c^{(1)}}{2}+1,\frac{N_c^{(2)}}{2}}2~cos(j~K_2)
  \Sigma^{(N_c^{(2)})}_{\lambda}(x=j,i\omega_n)
\end{split}
\end{equation}
\noindent where only diagonal parts of the Fourier transformed
self-energy are considered.

J. Hague {\it et.\ al.}~in ref.~\onlinecite{hague} considered the
following, alternative momentum-space {\it Ansatz} to combine the
different length-scales:

\begin{equation}
\label{eq:smallansatz}
\begin{split}
  \Sigma ^{(N_c^{(1)})}(K_1,i\omega _n ) &= \Sigma _{QMC}^{(N_c^{(1)})} (K_1,i\omega _n )\\
  & - \Sigma_{\lambda}^{(N_c^{(1)})} (K_1,i\omega _n ) + \bar \Sigma
  _{\lambda}^{(N_c^{(2)})} (K_1,i\omega _n )
\end{split}
\end{equation}

\begin{equation}
\label{eq:largeansatz}
\begin{split}
  \Sigma ^{(N_c^{(2)})} (K_2,i\omega _n )& = \Sigma _{\lambda}^{(N_c^{(2)})}(K_2,i\omega _n )\\
  &+ \bar \Sigma _{QMC}^{(N_c^{(1)})} (K_2,i\omega _n ) - \bar \Sigma
  _{\lambda}^{(N_c^{(1)})}(K_2,i\omega _n )
\end{split}
\end{equation}
\noindent where $\Sigma _{\lambda}^{(N_c^{(1)})}$ is the self-energy
obtained in the $\lambda$-approximation when implemented on the small
cluster. The self-energies $\Sigma^{(N_c^{(1)})}$ and
$\Sigma^{(N_c^{(2)})}$ on the small and large cluster, respectively,
exist on different grid sizes, and it becomes necessary to convert
self-energies from one to the other. This conversion is denoted by a
bar over the self-energy which denotes an interpolation when going
from a coarser grid to a finer one and a coarse-graining step
otherwise. For example, the large cluster self-energy
$\Sigma^{(N_c^{(2)})}$ is constructed from the explicit
$\lambda$-approximated self-energy on the large cluster, and two
interpolated small cluster (denoted by the superscript $(N_c^{(1)})$)
self-energies $\bar \Sigma _{QMC}^{(N_c^{(1)})}$ and $\bar \Sigma
_{\lambda}^{(N_c^{(1)})}$.  It is important to note that the coarse
graining in going from large- to small-cluster self-energies is not an
averaging over electron propagators but the term is used in this
context to imply an averaging of the self-energy within a
cluster-cell.

While the real space implementation of the {\it Ansatz} was a straight
forward combination of different length-scale elements, the momentum
implementation interpretation is more involved.  The $\lambda$-method
provides an estimate for the set of self-energy diagrams which convey
the long-ranged correlations of the system. However, since in
reciprocal space there isn't an explicit separation of length-scales,
the remaining short length-scale diagrams which are to be supplied by
the QMC calculation have to be identified. This can be accomplished by
removing the sub-set of $\lambda$-approximated self-energy diagrams on
the small cluster from the complete set of the QMC calculation hence
avoiding a double counting of the corresponding self-energy
contributions.

Within the traditional form of the DCA, all self-energies are
inherently causal for each individual cluster. One consequence of
causality is that $-\frac{1}{\pi}Im\Sigma(k,\omega)>0$.  In this {\it
  Ansatz} based implementation of the MSMB method however, causality
is not inherently guaranteed. The combination of self-energy
contributions of various length-scales in the {\it Ansatz} is only
assured to yield a causal multi-scale self-energy in the limit of the
cluster sizes approaching one another. Therefore, causality in these
schemes cannot be guaranteed and hence has to be monitored closely
throughout. For a more detailed discussion of this and the entire
momentum-space {\it Ansatz} see Ref.~\onlinecite{hague}.

\begin{figure}
\centerline{\includegraphics*[width=2.8in]{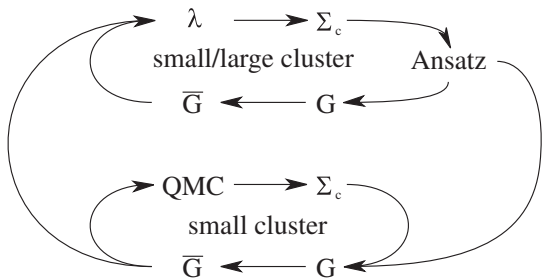}}
\caption{Flow chart for the {\it Ansatz} self-consistent implementation of the MSMB/DCA.}
\label{fig:Ansatz}
\end{figure}

The remaining self-consistent implementation of the multi scale method
(i.e.\ two cluster DCA) is similar to that of the traditional, single
cluster DCA.  Fig.~\ref{fig:Ansatz} depicts a flow chart of the
implementation of both {\it Ans\"{a}tze} in the scheme of the overall
self-consistency loop. The bottom loop shows the already discussed DCA
self-consistency loop using the QMC as a small cluster solver. For the
large cluster solver ($\lambda$-method) the self-consistency is
similar. In the overall scheme of the {\it Ansatz}, the two cluster
problems are combined in a fully self-consistent approach: After each
iteration of the QMC/DCA loop, the corresponding $\lambda$/DCA
contribution on the large cluster is evaluated.  The {\it Ansatz} is
used after each step to calculate new estimates for the self-energies
on both the small and large cluster, thus yielding a fully
self-consistent solution. The corresponding paths are shown in
Fig.~\ref{fig:Ansatz}. The converged {\it Ansatz}
self-energy will be dominated by the small cluster QMC self-energy
which contributes the strongest, short-ranged correlations.  The
remaining weaker long-ranged correlations are incorporated in the
difference of $\lambda$-approximated self-energies between the two
clusters.  In moving away from a fully self-consistent approach, the
self-consistency restrictions may be lessened to various degrees. Some
possible implementations will be discussed in further detail in
section~\ref{techasp}.

\section{Approximations to the vertex}
\label{lcs}

In this section we provide the remaining details of the
$\lambda$-approximation.  Previous studies, as well as our own
results, have shown that in the positive $U$ Hubbard model the
long-ranged contributions to the self-energy are dominated by the spin
and charge fluctuations of the system. In contrast, pairing
fluctuations are less significant and do not change the qualitative
multi-scale results, unless very low temperatures are considered.  The
self-energy determination in the results section for the lowest order
in $U$ approximation to $\Gamma^{\lambda}$ considers all three
channels. However, the remaining higher order cases are restricted to
only the self-energy contribution of the particle-hole channel. An
equivalent derivation to the following can be trivially extended to
the particle-particle channel.

Starting with the $\lambda$-approximation introduced in
sections~\ref{multiscale} (Eqs.~\ref{eq:gclust} and {\ref{eq:gamma}),
  the Bethe-Salpeter equation (Eq.~\ref{eq:feqg}) for the approximated
  large cluster reducible vertex function can be coarse-grained to the
  small cluster (see section~\ref{outlook} for further details):

\begin{equation}
\label{eq:flam}
\begin{split}
  \overline{F}^\lambda(K_1,K'_1&,Q_2;\imath \omega_n,\imath \omega_n',
  \imath \nu) =
  \Gamma^{\lambda}(K_1,K'_1; \imath \omega_n, \imath \omega_n')\\
  &+ \Gamma^{\lambda}(K_1,K''_1; \imath \omega_n, \imath \omega_n'')
  \times \overline{\chi}^0(K''_1,Q_2; \imath \omega _n,
  \imath \nu)\\
  &\times \overline{F}^{\lambda}(K''_1,K'_1,Q_2;\imath
  \omega_n'',\imath \omega_n', \imath \nu)
\end{split}
\end{equation}

\noindent where we coarse grained the bare particle-hole
susceptibility bubble (internal legs in Fig.~\ref{fig:bse} which are
lattice Green's functions).  We call this coarse grained
susceptibility $\overline{\chi}^0$

\begin{equation}
\begin{split}
  \overline{\chi}^0(K''_1,Q_2;\imath \omega _n, \imath \nu)= &-\frac{T
    N_c}{N} \sum_{\tilde{k}'',\imath \omega n}
  G(K''_1+\tilde{k}'',\imath \omega_n'')\\
  &\times G(K''_1+\tilde{k}''+Q_2,\imath \omega_n''+\imath \nu)~.
\end{split}
\end{equation}

\noindent Furthermore, while the coarse-graining (sum over internal
moment $\tilde{k}''$) is to the small cluster, the self-energy in $G$
originates from the large cluster. The resulting reducible vertex
function $\overline{F}^\lambda$ however, is still only defined for
small cluster momenta but incorporates large cluster corrections from
the coarse grained susceptibility $\overline{\chi}^0$. Due to these
long-ranged contributions, the reducible vertex constructed in this
manner is not equivalent to a QMC evaluated small cluster $F$.

The resulting Bethe-Salpeter equation~(\ref{eq:flam}) is most easily
solved for $\overline{F}^\lambda$ in the following matrix form

\begin{equation}
  \overline{F}^{(\lambda) -1}(Q_2,\imath \nu)=\Gamma^{(\lambda)-1}-\overline
  \chi^{0}(Q_2,\imath \nu)
\end{equation}

\noindent where the matrix indices correspond to the internal momenta
and frequency.

At this point, the self-energies can now finally be evaluated using
the Dyson equation as illustrated in Fig.~\ref{fig:Fig2}. On the large
cluster this yields:

\begin{equation}
\begin{split}
  &\Sigma^{(N_c^{(2)})}_{\lambda}(K_2,\imath \omega_n)=\\ 
&\frac{U T^2}{N_c^{(1)} N_c^{(2)}} \sum_{K'_1, Q_2, \imath \omega_n',\imath
    \nu }G(K_2+Q_2,\imath \omega_n+\imath \nu)\\
  & \overline{\chi}^0(K'_1,Q_2;\imath \omega _n', \imath \nu) \times (
  \overline{F}^{\lambda+-}(K'_1,M(K_2),Q_2;\imath \omega_n',\imath \omega_n, \imath \nu)\\
  &-\overline{F}^{\lambda\uparrow\downarrow}(K'_1,M(K_2),Q_2;\imath
  \omega_n',\imath \omega_n, \imath \nu)-U )~.
\end{split}
\end{equation}

\noindent Here we interpolate the small cluster momentum $M(K_2)\rightarrow K_2$  
(for details see section~\ref{outlook}) and
subtracted $U$ in the parenthesis to prevent an over
counting of the second order term. 

In the calculation of the transverse spin fluctuation part, recall
that $2\chi^{\pm} = \chi^{zz}$ where we define $\chi^\pm$ as the
correlation function of $\sigma^+$ and $\sigma^-$, and $\chi^{zz}$ as
the correlation function formed from $\sigma^z$. Then as $\chi^{zz} =
2(\chi^{\uparrow\uparrow} - \chi^{\downarrow\uparrow})$, we have that
$\chi^\pm = \chi^{\uparrow\uparrow} - \chi^{\downarrow\uparrow}$ and
$F^\pm=F^{\uparrow\uparrow} - F^{\downarrow\uparrow}$. This means that
for the self-energy on the large cluster,

\begin{equation}
\label{eq:siglam}
\begin{split}
  &\Sigma^{(N_c^{(2)})}_{\lambda}(K_2,\imath \omega_n)=\\
  &-\frac{U T^2}{N_c^{(1)} N_c^{(2)}} \sum_{K'_1, Q_2, \imath
    \omega_n',\imath \nu }
  G(K_2+Q_2,\imath \omega_n+\imath \nu) \\
  &\times \left(
    2\overline{F}^{\lambda\uparrow\downarrow}(K'_1,M(K_2),Q_2;\imath
    \omega_n',\imath \omega_n, \imath \nu)+U
  \right)\overline{\chi}^0(K'_1,Q_2;\imath \omega_n', \imath \nu)~.
\end{split}
\end{equation}

\noindent While for the real-space implementation of the {\it Ansatz},
knowledge of the large cluster self-energy in the
$\lambda$-approximation is sufficient, the momentum-space version
requires the corresponding self-energy diagrams on the small cluster
as well. An equivalent calculation on the small cluster yields for the
self-energy in $\lambda$-approximation

\begin{equation}
\begin{split}
  &\Sigma^{(N_c^{(1)})}_{\lambda}(K_1,\imath \omega_n)=\\
  &-\frac{U T^2}{{N_c^{(1)}}^2} \sum_{K'_1, Q_1, \imath
    \omega_n',\imath \nu }
  G_{c}(K_1+Q_1,\imath \omega_n+\imath \nu) \\
  &\times \left(
    2F^{\lambda\uparrow\downarrow}_{c}(K'_1,K_1,Q_1;\imath
    \omega_n',\imath \omega_n, \imath \nu) +U
  \right)\chi^0_{c}(K'_1,Q_1;\imath \omega _n', \imath \nu)
\end{split}
\end{equation}

\noindent where all single-particle propagators have been replaced by
$G_c$, including the ones entering the bare bubble $\chi^0$.

We want to reiterate that the $\lambda$-approximated self energy thus
obtained, is only accurate for long-ranged correlations. In order to obtain a
viable multi-scale solution, the neglected short-ranged correlations
have to be accounted for by means of the {\it Ansatz}, as outlined in
the previous section.

\subsection{First Order in $U$}
\label{fou}

The introduced non-perturbative MSMB method requires a detailed
knowledge of the small (QMC) cluster irreducible vertex. This
evaluation of $\Gamma$ however, poses a difficult and numerically
involved problem. We shall therefore initially only consider
perturbative approximations to the vertex within the MSMB method. In
approximating the irreducible vertex by the frequency independent
first order contribution i.e.\ $\Gamma(K,K';i\omega_n,i\omega_n')=U$,
the self-energy diagrams of Fig.~\ref{fig:Fig2} reduce to the well
known FLEX\cite{bickers,bickers1,bickers2} diagrams. In the
particle-hole channel, the corresponding self-energy is given by

\begin{equation}
  \Sigma^{(ph)}(k,\omega_n)=\frac{U T}{N}\sum_q \sum_m V^{(ph)}(q,\omega_m)
  G(k-q,\omega_n-\omega_m)~~.
\end{equation}                                                                                    
                         
\noindent The FLEX potential is defined by
                    
\begin{equation}
\label{eq:FLEX}
\begin{split}
  V^{(ph)} &=\chi^0_{ph}(q,\omega_n)-\frac{1}{2}
  \frac{\chi_{ph}^{0~2}(q,\omega_n)}{1+\chi^0_{ph}(q,\omega_n)} \\
  &+ \frac{3}{2}
  \frac{\chi_{ph}^{0~2}(q,\omega_n)}{1-\chi^0_{ph}(q,\omega_n)}
\end{split}
\end{equation}

\noindent where
                   
\begin{equation}
  \chi^0_{ph}(q,\omega_n)=-\frac{T}{N} \sum _{k} \sum_{\omega}
  G(k,\omega_m)G(k+q,\omega_m+\omega_n)~.
\end{equation}

Using the FLEX to address the long length-scale problem within a
multi-scale method was the scope of the work by J. Hague\cite{hague}.
Initially we will return to this simple cluster solver which is known to
provide qualitatively correct long length-scale properties.  We will
use the FLEX to illustrate some of the properties and problems
associated with the {\it Ansatz} in momentum space (see
Eqs.~\ref{eq:smallansatz} and ~\ref{eq:largeansatz}). After exploring
some of the encountered limitations, we will introduce variations of
the original implementation which expand the scope of FLEX
applicability within the MSMB scheme.

\subsection{Second Order in $U$}

At lower temperature or for larger $U$, higher-order terms in the
vertex are important.  The second order corrections to the irreducible
vertex function in the $\lambda$-approximation (i.e.~zero external
momentum transfer $q=0$ and frequency $\nu_n=0$) are shown in
Fig.~\ref{fig:2ndorder}.  The irreducible vertex is given in the spin
channel by

\begin{figure}
\centerline{\includegraphics*[width=3.3in]{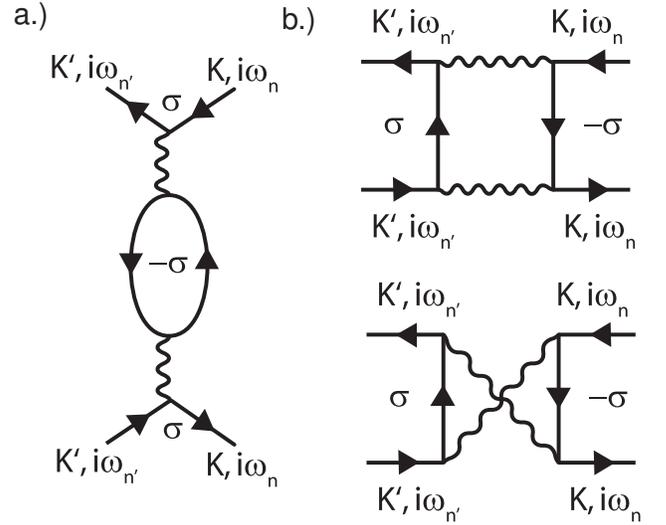}}
\caption{Second order diagrams for the vertex functions
  $\Gamma^2_{\uparrow\uparrow}$ - a.) and
  $\Gamma^2_{\uparrow\downarrow}$ - b.) for an external momentum
  transfer $q=0$.}
\label{fig:2ndorder}
\end{figure}

\begin{equation}
\begin{split}
  \Gamma^{(\lambda)s}(i\omega_n,i\omega_{n'};K_1&,K_1')= - U
  -\frac{U^2T}{N_c^{(1)}}\sum_{i\omega_{n''},K_1''} G(i\omega_{n''},K_1'')\\
  &\times G(i\omega_n+i\omega_{n'}-i\omega_{n''},K_1+K_1'-K_1'')
\end{split}
\end{equation}

\noindent and in the charge channel

\begin{equation}
\begin{split}
\Gamma^{(\lambda)c}(i\omega_n,i\omega_{n'}&;K_1,K_1')= + U 
+\frac{U^2T}{N_c^{(1)}}\sum_{i\omega_{n''},K_1''} G(i\omega_{n''},K_1'')\\
&\times (G(i\omega_n+i\omega_{n'}-i\omega_{n''},K_1+K_1'-K_1'') \\ 
&+2~G(i\omega_n-i\omega_{n'}+i\omega_{n''},K_1-K_1'+K_1''))~.
\end{split}
\end{equation}

\subsection{Full QMC vertex}

In the non-perturbative MSMB approach the full irreducible vertex, as
obtained by the QMC, is considered.  Within the QMC algorithm we are
unable to determine $\Gamma$  directly and instead calculate the
two-particle correlation function $\chi$. The irreducible vertex
function can be found by inverting the Bethe-Salpeter equation.  In
the spin-channel we have the two particle correlation function

\begin{equation}
\chi^s =\chi^0 +\chi^0 \Gamma^s\chi^s
\end{equation}

\noindent which is a matrix with elements both in frequency and
momentum (the same holds for the charge channel). The irreducible spin
vertex is denoted by $\Gamma^s$ and $\chi^0$ is the bare spin
susceptibility respectively.  The correlation functions are evaluated
in the QMC by sampling over the configuration space
(Hirsch-Hubbard-Stratonovich fields\cite{dca}) in one of two ways,
each one posing its own challenges.

One possibility is directly evaluating
$\chi(i\omega_n,k;i\omega_{n'},k')$ in frequency space.  This requires
the individual QMC Green's function to be Fourier transformed (FT)
from the time domain. However, since the calculation is limited to
finite time intervals $\Delta\tau$ the FT will incur substantial
high-frequency artifacts. In theory this can be improved by means of a
high-frequency conditioning, but no analytic form is available within
the QMC calculation to facilitate such a conditioning. This resulting
substantial artifacts are propagated into the irreducible vertex
function.

Alternatively $\chi$ can be evaluated in the time domain and only
Fourier transformed once the QMC averaging is complete. This results
in an accurate measurement of the two-particle correlation function by
the QMC but is significantly more computationally expensive.  Although
this approach provides an accurate measure for
$\chi(\tau_1,\tau_2;\tau_3,\tau_4)$, the Bethe-Salpeter equation can not
be be used to solve for $\Gamma$ in the time domain and hence the QMC averaged
$\chi$ has to be Fourier transformed first.  This once again results
in similar problems associated with the FT.  In the high temperature
regime where higher order corrections in $U$ are not important a
perturbative approach to high-frequency conditioning of the FT is very
successful.  However, as the temperature is lowered and the
significance of higher order corrections within the vertex grows, this
conditioning results in even larger artifacts. Therefore we use the 
frequency domain in our determination of $\chi$.

\section{Technical Aspects}
\label{techasp}

A MSMB approach based on a first order approximated vertex is
similar to the limited FLEX-hybrid approach previously considered by J.~Hague
{\it et.~al.}~\cite{hague}. In their work the intermediate
length-regime was addressed using the FLEX which was incorporated
within a multi-scale approach using the fully self-consistent momentum
space {\it Ansatz} (see Eqs.~\ref{eq:smallansatz} and
\ref{eq:largeansatz}). However, in the regime of stronger couplings
and/or lower temperatures, where a significant contribution of longer
ranged correlations is to be expected, the method fails. In this
section we propose changes to the implementation of the multi-scales
method which significantly improves its range of applicability by 1)
restriction of the {\it Ansatz} to the large cluster, 2) modification
of the cluster conversion, and 3) removal of the self-consist
implementation of the {\it Ansatz} on the large cluster.

\begin{figure}
\centerline{\includegraphics*[width=3.3in]{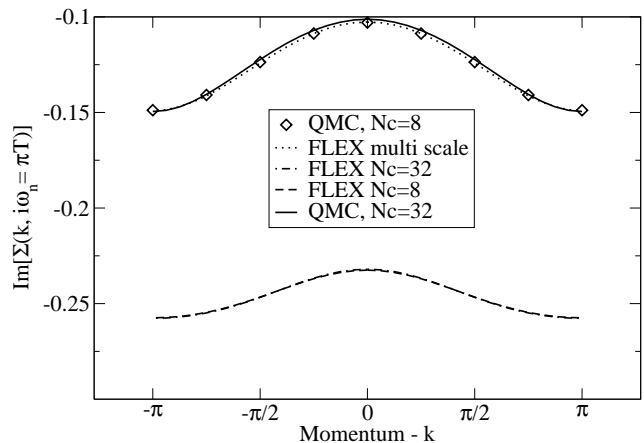}}
\caption{Imaginary part of the self-energy at lowest Matsubara frequency as obtained by various
cluster solvers and FLEX/MSMB method at $\beta=8$, $U=W=1.0$, and $n=0.75$. 
The MSMB cluster sizes are $N_c^{(1)}=8$ and $N_c^{(2)}=32$.}
\label{fig:compU02}
\end{figure}

In the original inception\cite{hague}, the {\it Ansatz} is evaluated
fully self-consistently within both the QMC and FLEX as discussed in
section \ref{ansatzsec}. The self-energy contributions of these two
clusters solvers are linked by the momentum space {\it Ansatz}.
However, long ranged-correlations, as are described by the large
cluster FLEX, are assumed weak. Hence, their presence in the effective
medium of the small cluster only minutely effects the QMC self-energy.
We thus neglect the {\it Ansatz} on the small cluster
(Eq.~\ref{eq:smallansatz}).  
With this modification, the small cluster problem can be solved
independently of the large cluster problem. This results in a
significant reduction in numerical complexity.

With the removal of the small cluster {\it Ansatz} condition, only a
unidirectionally cluster conversion of small cluster self-energies to
the large cluster remains. In this work,
a periodic cubic spline interpolation~\footnote{We omit the additional
  coarse graining of the self-energy in the original work. This is
  deemed appropriate since the interpolated small cluster self-energy
  wont recoup the lattice self-energy and hence a coarse graining to a
  cluster is merit-less.} is employed. This provides a good
approximation for the multi-scale self-energy in the high
temperature/small $U$ limit. Fig.~\ref{fig:compU02} shows the
imaginary part of the exact, QMC calculated and FLEX self-energies on
both clusters calculated at high $T$. In this regime, correlations are
short ranged in nature and thus well described by the small cluster
itself, while remaining long-ranged features, which are provided by
the large cluster FLEX, are insignificant. The interpolated small
cluster self-energy accurately replicates that of the explicit large
single cluster QMC result. This results in a vanishing FLEX contribution
within the {\it Ansatz} as the large and interpolated small cluster
FLEX results are identical and thus cancel each other (see
superimposed, lower set of curves in Fig.~\ref{fig:compU02}).

\begin{figure}
\centerline{\includegraphics*[width=3.3in]{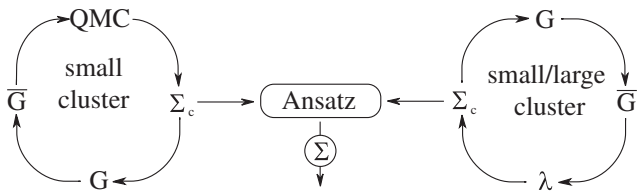}}
\caption{Flow chart depicting two independent self-consistent DCA calculations combined 
via an {\it Ansatz} to construct a MSMB self-energy.}
\label{fig:FAnsatz}
\end{figure}

In this high temperature regime (e.g.~Fig.~\ref{fig:compU02}) the fully self-consistent 
{\it Ansatz} as well as our modified one 
remain numerically stable. However, one inherent limitation of the
approach is already apparent. The observed magnitude of the negative
imaginary part of the QMC self-energy is significantly smaller than
the FLEX result. This overestimation of the self-energy by the FLEX deems
a first order approximation to the vertex to be insufficient.
A self-consistent implementation of the 
{\it Ansatz} aggregates the problem as the {\it Ansatz} self-energy 
used to initialize each FLEX iteration is
similar in magnitude to the QMC self-energy. This results in an
insufficient damping of the FLEX potentials and the subsequent overestimation
of the FLEX self-energy renders the
large cluster self-energy calculation numerically
unstable. This is an inherent problem in the self-consistent approach.
In an attempt to enhance the numeric stability of the MSMB method we
remove the self-consistent implementation of the {\it Ansatz} on the
large cluster.  This leaves two independent DCA calculations for the
small and large cluster which, after the individual problems are
converged, are combined by the {\it Ansatz}. The resulting
self-consistency scheme is depicted in Fig.~\ref{fig:FAnsatz}.  
However, the effective media
embedding the two cluster problems now are unaware of correlations as
determined by the other cluster solver technique. Therefore, the
effective medium of the large cluster now lacks the explicit short
ranged correlations as provided by the QMC. Similarly, the effective
medium of the small cluster only contains long-ranged correlations as
provided by the dynamical mean-field approximation.

The third modification enters in the determination of the small
cluster FLEX self-energy $\Sigma _{FLEX}^{(N_c^{(1)})}$. Rather than
calculating it explicitly, we obtain it by coarse graining the large
cluster self-energy onto the small cluster. The combination of all
changes introduced up to this point significantly increase the range
of low temperatures which can be addressed by a perturbative
MSMB approach.  

Throughout the remainder of the paper
we will restrict ourselves to the
non-self-consistent momentum space {\it Ansatz} as outlined in this
section. It indeed succeeds in addressing some of the problems
encountered in the original implementation and yields a numerically
stable MSMB method.

\section{Results}
\label{results}
Throughout the remainder of the paper, we evaluate the quality of the
MSMB method by comparing the self-energy to that of a 32 site, single
cluster DCA/QMC calculation.

\subsection{First Order in $U$}
\label{flex}

We begin the evaluation of the MSMB method by considering the first
order approximation to the irreducible vertex. The previous
section introduced a numerically stable approach to the perturbative 
MSMB method. This allows for the FLEX based MSMB treatment of 
stronger coupling/lower temperature regimes.

The major limitation of this method however remains in the
still significant difference in magnitude of the QMC and FLEX
self-energies leading to an overestimation of long length-scale
features introduced by the FLEX. In Fig.~\ref{fig:compU10} it is quite
apparent that at lower temperatures the FLEX MSMB implementation
overestimates the size of the long length-scale features i.e.\
amplitude of the oscillations in the imaginary part of the 
self-energy. This is yet further 
indication that a bare approximation to the vertex is inadequate
to address the intermediate length regime. 

\begin{figure}
\centerline{\includegraphics*[width=3.3in]{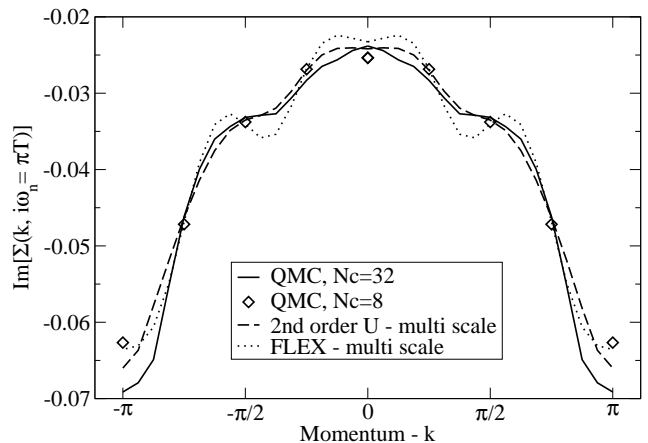}}
\caption{Imaginary part of the self-energy at lowest Matsubara
  frequency as obtained by the MSMB method using first (FLEX) and
  second order approximated irreducible vertices $\Gamma$ in
  comparison to the large single-cluster QMC results at $\beta=31$,
  $U=W=1.0$, and $n=0.75$. Multi-scale results are for cluster sizes
  $N_c^{(1)}=8$ and $N_c^{(2)}=32$.}
\label{fig:compU10}
\end{figure}

\subsection{Second order in $U$}
\label{2orderU}

We move to include second order corrections in the
vertex which is expected to mitigating the effects of the underestimation
to the vertex in the FLEX. This lessens the difference in magnitude between the QMC
and $\lambda$ self-energy within the {\it Ansatz} (as observed in Fig.~\ref{fig:compU02}) and
thus increase the applicability of the MSMB method to lower
temperatures.    

Fig.~\ref{fig:compU10} compares the imaginary part of the self-energy
at lowest Matsubara frequency for various degrees of approximation.
In contrast to the gross overestimation of the first order
approximation to $\Gamma$, the inclusion of second order in $U$
corrections within the MSMB method successfully captures the long
length-scale features of the large single-cluster QMC self-energy
throughout most of the Brillouin zone.  The largest deviations in the
multi-scale self-energy are found about the corners of the Brillouin
zone i.e.\ $k=\pm \pi$. Within the {\it Ansatz}, the difference of the
small and large cluster $\lambda$ self-energy provides the long
length-scale features thus partially restoring the self-energy
information which was lost by coarse graining to the small QMC
cluster. However, using a $\lambda$ self-energy with a negative
imaginary part which is considerably larger than that of the small
cluster QMC result in an overestimation of this correction,
predominately in regions encompassed by steep gradients in the self-energy.

Fig.~\ref{fig:methods} further illustrates the pathology of the {\it
  Ansatz} associated with the difference in magnitude of the cluster
self-energies.
We show the imaginary part of the self-energy as obtained by the
various $\lambda$-approximations on the large cluster in comparison to
the large single-cluster QMC result.
The magnitude of the negative imaginary part of the self-consistent
FLEX (first order in $U$ approximation) self-energy is, as was
previously indicated, very large compared to the magnitude of the QMC
result. The new approach of including second order corrections in $U$
in the vertex function, succeeds in yielding a self-energy which
resembles that of the QMC more closely. Similarly to the numeric
instability encountered with the FLEX approximation, the self-consistent 
determination of the $\lambda$ self-energy using the large cluster 
{\it Ansatz} also
yields a larger self-energy but doesn't encounter the catastrophic
divergence of the FLEX. We find that a second order $\Gamma$ combined
with a non-self-consistent {\it Ansatz} provides the best multi-scale
solution to the problem.

\begin{figure}
\centerline{\includegraphics*[width=3.3in]{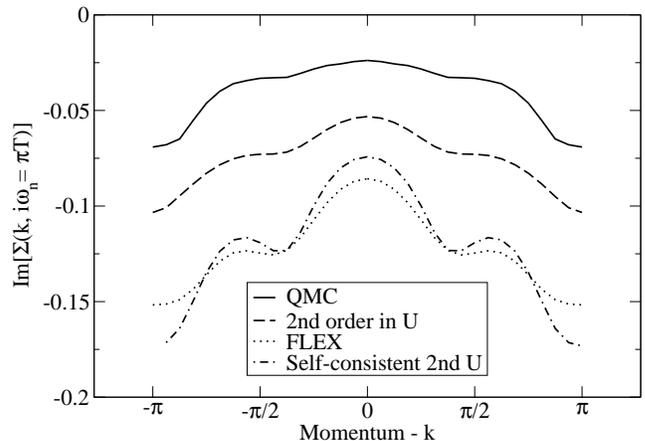}} 
\caption{Imaginary part of the $\lambda$-approximated self-energy at
  lowest Matsubara frequency for the large cluster ($N_c^{(2)}=32$)
  using various cluster solvers in comparison to the single cluster
  QMC result at $\beta=31$, $U=W=1.0$, and $n=0.75$. Also shown is the
  overestimated {\it Ansatz} self-consistent self-energy of the MSMB
  method for the second order approximated $\Gamma$.}
\label{fig:methods}
\end{figure}

Now that we have established the viability of the MSMB technique we
take a closer look at the dependence of the small cluster size on the
quality of the multi-scale result.  Within the MSMB method the QMC
constitutes the computationally most expensive part. Hence, we want to
restrict the calculations to the smallest possible QMC cluster
($N_c^{(1)}$) without any significant loss of quality in the
multi-scale results. It is therefore important to study the dependence
of the multi-scale results on the size of the small cluster.

Fig.~\ref{fig:smallclust} shows the multi-scale self-energy for a
variety of small cluster sizes as obtained by the MSMB method using
the second order approximation to the $\lambda$ vertex. It is apparent
that a cluster size of $N_c^{(1)}=4$ is too small to adequately
capture the main $k$ dependence (overall magnitude) of the QMC's
self-energy.  Considering such a small QMC cluster grossly
misrepresents the range of the QMC self-energy
and is not recovered in the MSMB method. This is an indication that
correlations beyond the length-scales of the small cluster are still significant and not
sufficiently well approximated by the $\lambda$-method. A small
cluster size of $N_c^{(1)}=8$ on the other hand, appears adequate and
only slightly underestimates the self-energy in the vicinity of
$k=\pm\pi$.% where the gradient is steepest.  
In this area we continually observe significant remnants 
of the coarse graining of the
QMC which is inadequately restored by the MSMB method.

\begin{figure}
\centerline{\includegraphics*[width=3.3in]{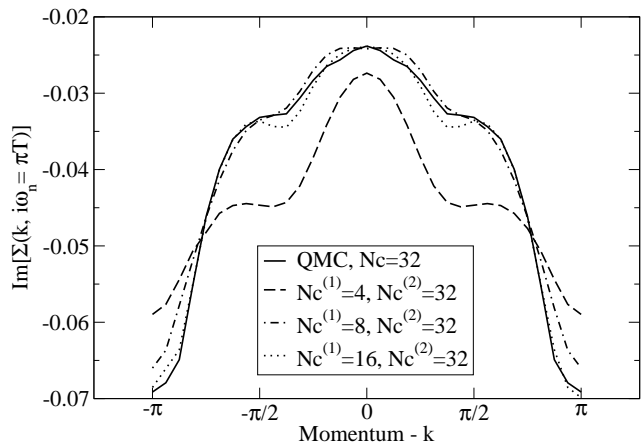}}
\caption{Imaginary part of the self-energy at lowest Matsubara
  frequency as obtained by the MSMB approach using the second order
  $\lambda$-approximation and large single-cluster QMC.  Results are
  for various small cluster sizes $N_c^{(1)}$ at $\beta=31$,
  $U=W=1.0$, and $n=0.75$.}
\label{fig:smallclust}
\end{figure}

Up to this point we have focused our investigation on the momentum
dependence of the self energy at small Matsubara frequency.  Since the
visual representation of the self-energy features at multiple, larger
Matsubara frequencies would be cumbersome, we proceed to further
illustrate the strengths of this MSMB technique by focusing on the
presence of spin-charge separation in the system which is manifest in
the full frequency dependent self-energy.

One dimensional systems have been shown to be non-Fermi liquids.
Amongst other unique features they are known to exhibit spin-charge
separation\cite{senechal,kim,preuss}. This very intriguing property
manifests itself by the complete decoupling of spin and charge degrees
of freedom. The single-particle spectra of such systems exhibits two
unique peaks corresponding to either spin or charge excitations which
move independently of each other. Due to the involved nature of
extracting spectra from the MSMB method, we are at this time unable to
directly identify the presence of any such feature in our results.
One characteristic of such a separation however, is the presence of
two distinct velocities corresponding to charge and spin respectively.
Following an approach by Zacher\cite{zacher}, we examine the MSMB
Matsubara frequency Green's function to the possible presence of
spin-charge separation.  This is done by fitting Green's function
obtained from the MSMB method to that of the Luttinger model solution

\begin{equation}
\label{eq:luttinger}
\begin{split}
  &G^{(LM)}_{v_1, v_2, \kappa _{\rho}}(x,\tau)=\\
  &\frac{{e^{ik_Fx}c}} {\sqrt{v_1 \tau + ix}\sqrt{v_2 \tau +
      ix}(x^2+v_2^2 \tau^2) ^{-(\kappa _ \rho +1/\kappa _ \rho-2)/8}}~~,
\end{split}
\end{equation}

\noindent where $v_1$ and $v_2$ are the spin and charge velocities and 
$c$ is a normalization constant.
We use the approximation for the correlation exponent $\kappa _{\rho}=1$ which is deemed
sufficient by Zacher for purposes of identifying the presence of two
different velocities. In order to accurately perform a fit with our
data we need to additionally coarse grain (see Eq.~\ref{coarg}) the
Luttinger liquid Green's function to obtain a fitting function in-line
with the idea of the DCA. This fit yields values for both the spin and
charge velocities when fitted at $k=\pi/2$ which is the DCA momentum
closest to the Fermi wave vector.  This specific choice of $k$ value
is motivated by fact that the Luttinger model solution is based on a
low energy approximation of the Hubbard model where a linearized
dispersion around the Fermi vector is assumed. For the parameters in
Fig.~\ref{fig:velfit} the Fermi wave vector ($k_F \approx 1.2$) falls
into the cluster cell about $k=\pi/2$.

\begin{figure}
\centerline{\includegraphics*[width=3.3in]{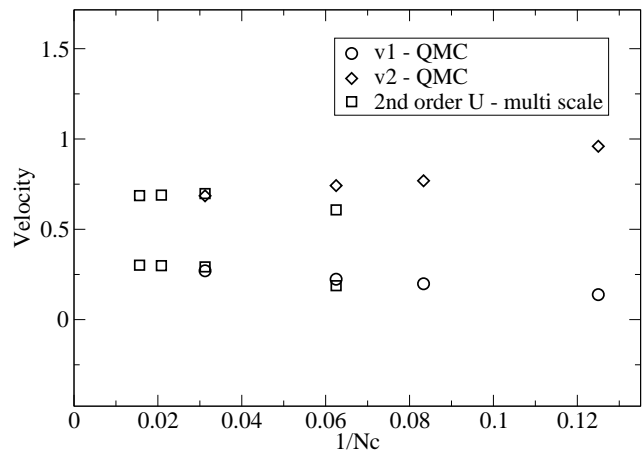}}
\caption{Spin and charge velocities ($v1$ and $v2$ respectively)  
  obtained by fitting the different results with the
  Luttinger Green's functions (Eq.~\ref{eq:luttinger}) about cluster
  momentum $k=\pi/2$ for $\beta=31$, $U=W=1.0$, and $n=0.75$.
  Multi-scale results are for a small cluster size $N_c^{(1)}=8$ and
  different large cluster sizes, and the QMC velocities were obtained
  from a single 32 site cluster calculation.}
\label{fig:velfit}
\end{figure}

Fig.~\ref{fig:velfit} shows the spin and charge velocities ($v_1$ and
$v_2$ respectively) obtained by using the single cluster QMC results
for smaller cluster sizes and multi-scale results for larger ones.
For intermediate cluster sizes, where both the QMC and the MSMB method
are feasible, we see a good match between the two methods. For larger
cluster sizes where the QMC approach becomes infeasible the
multi-scale results fall close to the QMC extrapolated values.
Furthermore, as we extrapolate the multi-scale results to the infinite
cluster size limit we find two different velocities: $v_1=0.311$ and
$v_2=0.674$. These values compare rather well to those obtained by
Zacher's {\it et.~al}.\cite{zacher} grand-canonical QMC calculation 
for a cluster size of 64, $\beta=80$
and all other parameters equal to ours: $v_1=0.293 \pm 0.019$ and
$v_2=0.513 \pm 0.023$.  Although the temperature in
Fig.~\ref{fig:velfit} is slightly higher compared to that of Zacher's
results, we observed only a small temperature dependence of the fitted
velocities in our calculations.  The similarity between the two
results are quite remarkable considering that our calculations were
based on a substantially smaller 8 site QMC cluster and hence less
subject to the sign-problem ~\footnote{The DCA inherently has a lesser
  sign problem within the QMC compared to finite size
  approaches~\cite{dca} and in the combination with a smaller cluster
  size results in a significantly larger average sign.}.  These are
yet further indications that the MSMB method indeed successfully
captures the long length-scale physics of the model and is in good
quantitative agreement with large single-cluster QMC calculations.

\subsection{Full QMC Vertex}
\label{lambda}

The previous two sections considered initially a first order
approximated irreducible vertex function (i.e.\ FLEX) and then
proceeded to show the significant improvement one can achieve by
including second order corrections to $\Gamma$ within the MSMB method.
Both of these perturbative approaches however resulted in an
overestimation of the long length-scale features of the self-energy at
low temperatures.  We now return to the non-perturbative
$\lambda$-approximation using the irreducible QMC vertex.  This
approximation is expected to significantly improve the results,
especially at lower temperatures where higher order corrections to the
bare vertex are important and measurably contribute to the
self-energy.

\begin{figure}
\centerline{\includegraphics*[width=3.3in]{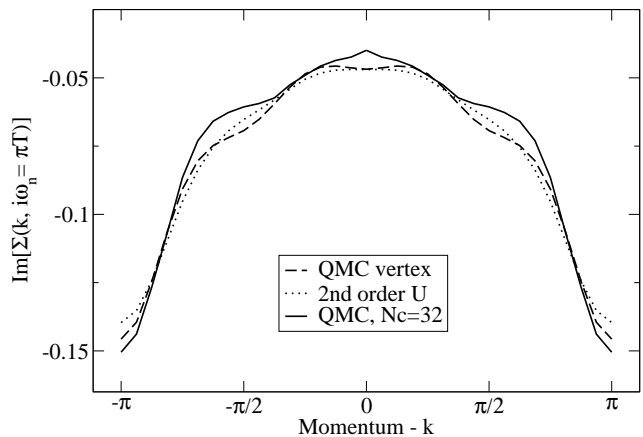}}
\caption{Imaginary part of the self-energy at lowest Matsubara
  frequency as obtained by the MSMB method using second order
  approximated and full QMC evaluated $\Gamma$s in comparison to a
  large single cluster QMC results at $\beta=31$, $U=\frac{3}{2}W=1.5$, and
  $n=0.75$. Multi-scale results are for cluster sizes $N_c^{(1)}=8$
  and $N_c^{(2)}=32$.}
\label{fig:vertexcomp}
\end{figure}

Fig.~\ref{fig:vertexcomp} compares the imaginary part of the
multi-scale self-energies obtained by considering both the second
order vertex and the full QMC vertex, to the numerically exact large
single-cluster QMC self-energy. While the resulting multi-scale
self-energy utilizing the full vertex function gives a better
approximation for some momenta, the second order vertex solution
remains preferable for others. This behavior may be somewhat
unexpected since the inclusion of higher order corrections in the
vertex was expected to further improve the overall quality of the
multi-scale solution. However, we believe that this discrepancy arises
from difficulties in extracting the exact vertex related to problems
with high frequency conditioning in the QMC and is not an intrinsic
problem of the method.
As a consequence, the consideration of second order corrections in the
$\lambda$-approximation remains the most useful at this time.  It
successfully allows for the exploration of lower temperatures - a
regime where the reducible vertex function develops a richness in features.

\subsection{Real Space Ansatz}
\label{rsansatz}

Up to this point, the various approximations to the irreducible vertex
function within the MSMB technique were implemented using the
momentum-space {\it Ansatz}. We now return to the previously
introduced implementation of the real-space {\it Ansatz} (see
Eq.~\ref{eq:realspace}). The significant difference is that the
separation of length-scales in the real-space approach is
straightforward, and therefore over-counting of diagrams is not an
issue. The second order vertex approach was shown to be widely
successful and will hence be used to investigate this {\it Ansatz}
implementation as well. It should be noted that both versions of the
{\it Ansatz} are treated identically as far as the implementation of
self-consistence is concerned (see Sec.~\ref{flex}).

\begin{figure}
\centerline{\includegraphics*[width=3.3in]{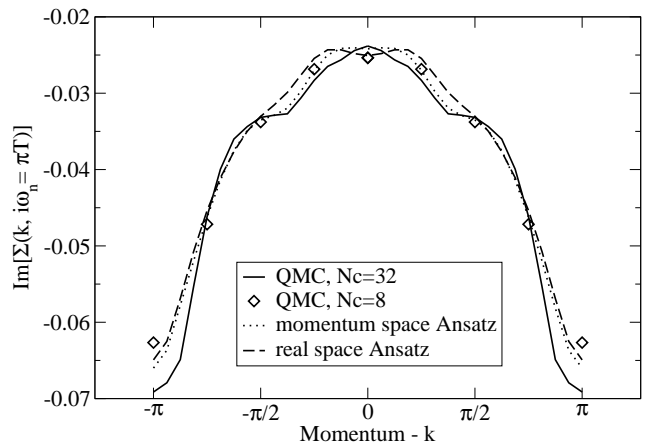}}
\caption{Imaginary part of the self-energy at lowest Matsubara
  frequency as obtained by the MSMB method using the second order
  approximated $\Gamma$ in conjunction with the real-space and
  momentum-space based {\it Ansatz} at $\beta=31$, $U=W=1.0$, and
  $n=0.75$.  Also shown is the large single-cluster QMC result.
  Multi-scale results are for cluster sizes $N_c^{(1)}=8$ and
  $N_c^{(2)}=32$.}
\label{fig:rsresult}
\end{figure}

Fig.~\ref{fig:rsresult} compares the multi-scale self-energy of the
two {\it Ansatz} implementations to the large single cluster QMC
result.  It is quite apparent that the real-space implementation of
the MSMB method is inferior. The momentum-space {\it Ansatz} provides
a self-energy more closely resembling the exact result throughout the
Brillouin zone.

The real-space {\it Ansatz} provides an intuitive and simple way to
combine the different length-scales of the problem in contrast to a
more complicated implementation in momentum space.  Although inferior,
the real-space {\it Ansatz} remains a viable, numerically stable
alternative.  It furthermore provides an alternative means of
interpolating the small cluster self energy by neglecting the large
cluster self-energy contributions in Eq.~\ref{eq:realspace}, which
were found to be negligible in this approach. The resulting
interpolated small cluster QMC self-energy (not shown) closely
resembles that of the the real-space {\it Ansatz} based MSMB method.

\section{Numerical Considerations and Outlook}
\label{outlook}

In the development of this MSMB method we were forced to employ
various approximations due to current computational limitations. The
largest numerical concession was made in section~\ref{lcs} where we
restricted the calculations of $F^{\lambda}$ to the small cluster (see
Eq.~\ref{eq:flam}). Ideally, the irreducible vertices $\Gamma$ would
be interpolated and the full reducible vertex evaluated on the large
cluster in turn. This calculation however would scale as
$(N_c^{(2)}N_l)^4$ ($N_l$ is the number of time-slices in the
QMC\cite{dca}) and hence provide little advantage over a single
cluster QMC calculation. We therefore restrict the evaluation of
$\bar{F}^{\lambda}\bar{\chi}^0$ in Eq.~\ref{eq:siglam} to the small
cluster and interpolate the product of $\chi^0$ and $F^\lambda$ to the
large cluster.

With the onset of peta-scale computing we will be able to make two
fundamental improvements to the MSMB approach in the near future.
Initially, we will gain the ability to include the fully momentum and
frequency dependent $\Gamma$ in our calculation, thus eliminating the
necessity of the $\lambda$-approximation (Eq.~\ref{eq:gamma}).
Inherent in this modification is an explicit account of the correct
short ranged physics hence removing the need of the {\it Ansatz}.
However, the memory and CPU requirements for this type of calculation
scale as $(N_c^{(1)}N_l)^3$.  For a large cluster with $N_c=16$ and
$N_l=100$, this would require 66G of double precision complex storage,
far exceeding the memory associated with a single CPU. These
staggering memory requirements can currently only be met by some
shared-memory parallel processing (SMP) super-computers.

In the second improvement, the approximation of the large cluster
$\Gamma$ by the small cluster QMC one can be replaced by one utilizing
the fully irreducible vertex $\Lambda$ (the vertex which is
two-particle irreducible in both the horizontal and vertical plane).
This results in a self-consistent renormalization of $\Gamma$ via the
Parquet equations~\cite{dominicis,bickers}, and hence an inclusion of long
ranged correlations in the crossing channel (see
Fig.~\ref{fig:parquet}) which are missing in both the
$\lambda$-approximation and the $\Gamma$-based approximation described
above.

\begin{figure}
\centerline{\includegraphics*[width=3.3in]{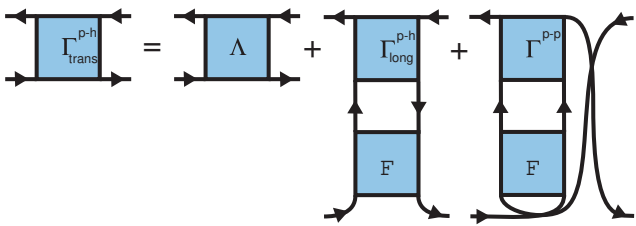}}
\caption{Parquet equation relating the transverse particle-hole irreducible vertex 
$\Gamma^{p-h}_{trans}$ to the fully irreducible vertex $\Lambda$ plus contributions from the
longitudinal and particle-particle cross channels. Similar relations 
apply for the remaining channels (not shown).}
\label{fig:parquet}
\end{figure}

\begin{figure}
\centerline{\includegraphics*[width=3.3in]{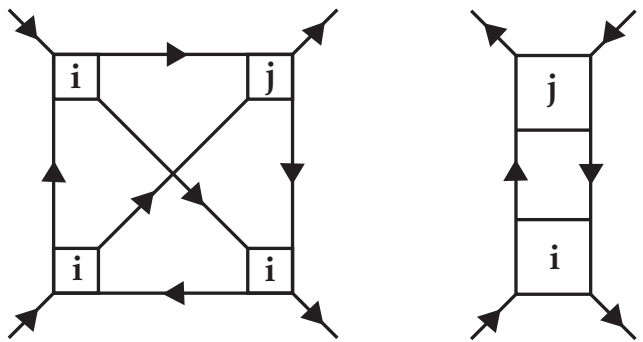}}
\caption{Lowest order non-local corrections to the fully-irreducible
  vertex $\Lambda$ (left) and the vertex $\Gamma$ (right).}
\label{fig:vertex}
\end{figure}

The superiority of the latter approach becomes clear in the
high-dimensional limit, where it is $\Lambda$, not $\Gamma$, which
becomes local. This can be shown by considering the simplest non-local
corrections to the respective vertices $\Lambda$ and $\Gamma$ in
Fig.~\ref{fig:vertex}. The boxes represent a set of graphs restricted
to site $i$ (local) and $j$ (neighboring) respectively. In the limit
of high dimensions, each site $i$ has $2D$ adjacent sites $j$. The
contributions of each leg within the vertex in the limit $D
\rightarrow \infty$ is $G(r) \sim D^{-r/2}$ (for details see
Ref.~\onlinecite{aryanpour}). This results in a contribution to the
correction of ${\cal O}(D^{-1})$ for the two legs in $\Gamma$ and
${\cal O}(D^{-3/2})$ for $\Lambda$.  Thus, the non-local corrections
to $\Lambda$ including all neighboring sites $j$ falls off as
$D^{-1/2}$ and becomes local in the infinite-dimensional limit.  In
contrast, the corrections to $\Gamma$ remain of order one. Therefore,
in the high-dimensional limit, $\Lambda$ is local while $\Gamma$ has
non-local corrections.  In finite dimensions, we would expect that
$\Lambda$ is more compact than $\Gamma$ whenever the single-particle
Green function falls quickly with distance. Then $\Lambda$ should
always be better approximated by a small cluster calculation than
$\Gamma$. (Despite the fact that $\Gamma$ has non-local corrections,
one can easily show that in the high-dimensional limit, all of the
methods discussed here will yield the same self energy and
susceptibilities since the non-local corrections to $\Gamma$ fall on a
set of zero measure points).

In employing the solution to the Parquet equation in a MSMB method we would 
be able to resolve two major limitations of the current approach: 1) An implementation 
considering the full frequency and momentum dependent vertex will be devoid of the 
causality problems associated with the self-energy mixing of the two cluster sizes. 
2) The approach constitutes a conserving approximation for the large cluster self-energy. 
Given these potential gains of a future method, we have to stress the extensive 
computational demands associated with this approach. While in a $\Gamma$ based 
implementation a trivial numerical parallelization of the problem leaves manageable 
demands, the complex nature of the Parquet approach requires substantial future 
development.

\section{Conclusion}
\label{conclusion}

We have introduced a numerically feasible MSMB extension to the DCA. In this method the 
lattice problem is mapped onto that of two embedded clusters, dividing the problem into three 
length-scales. Correlations on each of the length-scales are approximated commensurate with the
strength of the correlation on the respective scale.
The intermediate length regime, which bridges the explicit treatment of short ranged correlations
by means of the QMC to the long ranged dynamical mean-field one, 
is addressed in a diagrammatic long-wave 
length approximation based on the two particle irreducible vertex of the small cluster. The 
first order approximation to the vertex results in the FLEX but including higher order 
corrections result in a substantial better multi-scale result when compared to explicit large 
cluster QMC calculations. This can be attributed to the significance of higher order diagrams at 
lower temperatures. We proceeded to show that our MSMB results indicate spin and charge separation
and the obtained velocities compare favorably to significantly larger finite size QMC calculation.
The inclusion of the explicit QMC calculated vertex is currently still 
limited but further work in this direction looks promising. However, in any of the introduced 
implementations, the MSMB approach provides a means to adequately address large cluster 
problems on all length-scales at significant lower computational expense.

\section*{Acknowledgments}

We acknowledge useful conversations with D. Hess. This research was supported by grants NSF 
DMR-0312680, NSF DMR-0113574 and  NSF  SCI-9619020 through resources provided by the 
San Diego Supercomputer Center.  T. Maier and C. Slezak acknowledge support from the 
Center for Nanophase Materials Sciences, Oak Ridge National Laboratory, which is funded 
by the Division of Scientific User Facilities, U.S. Department of Energy.

During the completion of this paper we learned of two related studies Ref.~\onlinecite{toschi} 
and ~\onlinecite{kusunose} where long-ranged correlations are addressed in a
two-length scale, non-self-consistent
approach.

\end{document}